\documentclass[twocolumn,amsmath,amssymb,prl,superscriptaddress,floatfix]{revtex4}

\usepackage{graphicx}
\usepackage{dcolumn}
\usepackage{bm}

\begin{document}


\title{Dynamics of Dark-Bright Solitons in Cigar-Shaped Bose-Einstein Condensates}





\author{S. Middelkamp}
\affiliation{%
Zentrum f\"ur Optische Quantentechnologien,
Universit\"at Hamburg,
22761 Hamburg, Germany}

\author{J.J. Chang}
\affiliation{%
Washington State University,
Department of Physics \& Astronomy,
Pullman, WA 99164 USA}

\author{C. Hamner}
\affiliation{%
Washington State University,
Department of Physics \& Astronomy,
Pullman, WA 99164 USA}

\author{R. Carretero-Gonz\'{a}lez}
\affiliation{Nonlinear Physics Group,
Departamento de F\'{\i}sica Aplicada I,
Universidad de Sevilla,
41012 Sevilla, Spain}

\author{P. G.\ Kevrekidis }
\affiliation{Department of Mathematics and Statistics, University of Massachusetts,
Amherst MA 01003-4515, USA}

\author{V. Achilleos}
\affiliation{Department of Physics,
University of Athens,
Panepistimiopolis,
Zografos, Athens 157 84, Greece}

\author{D.J.\ Frantzeskakis}
\affiliation{Department of Physics,
University of Athens,
Panepistimiopolis,
Zografos, Athens 157 84, Greece}

\author{P. Schmelcher}
\affiliation{%
Zentrum f\"ur Optische Quantentechnologien,
Universit\"at Hamburg,
22761 Hamburg, Germany}

\author{P. Engels}
\affiliation{%
Washington State University,
Department of Physics \& Astronomy,
Pullman, WA 99164 USA}


\begin{abstract}\label{txt:abstract}
We explore the stability and dynamics of dark-bright solitons in two-component elongated Bose-Einstein condensates
by developing effective 1D vector equations as well as solving the corresponding 3D Gross-Pitaevskii equations. A strong dependence of the oscillation frequency and of the stability of the dark-bright (DB) soliton on the atom number of its components is found. Spontaneous symmetry breaking leads to
oscillatory dynamics in the transverse degrees of freedom for a large occupation of the component supporting the dark soliton.
Moreover, the interactions of two DB solitons are investigated with special
emphasis on the importance of their relative phases. Experimental results showcasing dark-bright soliton dynamics and collisions in a BEC consisting of two hyperfine states of $^{87}$Rb confined in an elongated optical dipole trap are presented.

\end{abstract}

\maketitle

{\it Introduction.} Multi-component systems of nonlinear waves are a
fascinating topic with a rich and diverse history
spanning a variety of areas, including Bose-Einstein condensates (BECs) in atomic
physics~\cite{BECBOOK}, optical fibers and crystals
in nonlinear optics~\cite{yuri}, and integrable systems in mathematical
physics~\cite{ablowitz}. Of particular interest are the
so-called ``symbiotic solitons'',
namely structures that would not
otherwise exist in one-component settings, but can be supported by
the interaction between the optical or atomic species components.
A prototypical example of such a structure is the
dark-bright (DB) soliton in self-defocusing, two-component systems,
whereby the dark soliton (density dip) which typically arises in
self-defocusing media~\cite{BECBOOK,yuri,ablowitz,djf} creates,
through nonlinearity, a trapping mechanism that localizes a
density hump (bright soliton) in the second component.

Dark-bright solitons were experimentally created in photorefractive
crystals
\cite{seg1},
while their interactions were partially monitored in~\cite{seg2}.
Upon realization of multi-component atomic
BECs~\cite{Myatt1997a,Hall1998a,Stamper-Kurn1998b}, it was
predicted that similar structures would exist
therein~\cite{buschanglin}. While theoretical developments
along
this direction were extended in even more complex
settings (such as the spinor system of Ref.~\cite{DDB}),
{\it stable} DB solitons were observed only recently in two-component
BECs \cite{hamburg}, leading to a
renewed interest in this area. Relevant recent works include
the interaction between
DB solitons~\cite{rajendran,berloff} and their
higher-dimensional generalizations \cite{VB}.

In order to study DB solitons in two-component elongated BECs we will use both
effectively one-dimensional (1D) mean-field models and the full 3D Gross-Pitaevskii equation (GPE).
For a single species, quasi-1D descriptions
rely on the non-polynomial Schr{\"o}dinger equation (NPSE)
\cite{sal1} and the Gerbier-Mu\~noz-Mateo-Delgado equation (GMDE) \cite{ger}; these models have been used
in studies of dark solitons in the dimensionality crossover between 1D and 3D,
yielding excellent quantitative agreement with experimental observations
\cite{ourstuff}. However, in multi-component settings only the NPSE equation has been derived~\cite{Salasnich06}.
In the present work, we first develop the GMDE for two-component BECs and then
investigate the DB soliton statics and dynamics
using the NPSE, the GMDE and the 3D GPE.
Varying the
atom number of either the dark ($N_D$) or the bright ($N_B$) component, we find that in a harmonic trap
{\it the soliton oscillation period may change by nearly
one
order of magnitude}; most notably, the bright component is shown to slow down
the oscillation of the dark one. Our investigation reveals
a feature absent in the
dark soliton dynamics in one-component BECs, namely
that {\it a single DB soliton may become dynamically unstable}.
Increasing $N_D$ and $N_B$ reveals
a deviation from the effective 1D description: specifically, {\it an
increase of $N_D$ leads to a  spontaneous breaking of the cylindrical
symmetry},
manifested in a transversal oscillation of the bright component, and a subsequent decrease of the axial DB oscillation frequency. Moreover, we analyze
the interaction between multi-DB solitons and the role of their relative phase.
Our results
pertain to the hyperfine states $|1,-1\rangle$ and $|2,0\rangle$ of $^{87}$Rb, as in
the experiment of Ref.~\cite{hamburg}, and also for the states $|1,-1\rangle$
and $|2,-2\rangle$ of $^{87}$Rb in an optical dipole trap. For the latter states, {\it we
present experimental results concerning
the DB soliton oscillations and
multi-DB soliton interactions} that further support our findings.

{\it Effective 1D Theory.} The macroscopic wave functions of Bose condensed
atoms in two different internal states obey the following vector
GPEs~\cite{BECBOOK}:
\begin{equation}
\imath \hbar \frac{\partial \psi_k}{\partial t} =
\Bigl( -\frac{\hbar^2\mathbf{\nabla}^2}{2M}
 + U
 +g_{kk}|\psi_k|^2
 +g_{12}|\psi_{3-k}|^2\Bigr)\psi_k ,
 \label{eq: coupled GP 3d}
\end{equation}
where $\psi_k(\mathbf{r})$ are the macroscopic wave functions
($k={1,2}$), normalized to $N_D$ and $N_B$
for the dark and bright soliton components, respectively,
$g_{ij}=4\pi\hbar^2a_{ij}/M$ are the effective nonlinear coefficients
due to the $s$-wave scattering for $i,j=1,2$, and
$U(\mathbf{r})$ is the confining potential.
For a highly anisotropic trap, we first factorize
the wave function as
$\psi_k(\mathbf{r},t)=\phi_k(\mathbf{r_{\perp}};x)f_k(x,t)
$ \cite{sal1,ger},
we substitute in Eq.~(\ref{eq: coupled GP 3d}), multiply
by $\phi^{\star}_k(\mathbf{r_{\perp}};x)$ and, finally,
integrate over the transverse directions; this leads to the following effective 1D model:
\begin{equation}
\left[\imath \hbar \frac{\partial}{\partial t} +
 \frac{\hbar^2}{2M}\frac{\partial^2}{\partial x^2} -V(x) \right] f_k
 =\mu_{\perp k}[f_k] f_k,
 \label{eq: f}
\end{equation}
where $V(x)=M\omega_{x}^2 x^2/2$ is the axial potential and the transverse
chemical potential $\mu_{\perp k}=\mu_{\perp k}[f_k(x,t)]$ is a functional of $f_k$:
\begin{eqnarray}
\mu_{\perp k}[f_k]&=& \int d^2 \mathbf{r}_{\perp} \phi_k^{\star} \Bigl( -\frac{\hbar^2}{2M}\mathbf{\nabla_{\perp}}^2
+ \frac{1}{2}M\omega_{\perp}^2 r_{\perp}^2
\nonumber\\
& +&g_{kk}|\phi_k|^2|f_k|^2 +g_{12}|\phi_{3-k}|^2|f_{3-k}|^2  \Bigr)\phi_k, \label{eq: muperp k}
\end{eqnarray}
where $(\omega_{x},\omega_{\perp})$ are the trap frequencies along the
longitudinal (axial) and transverse directions, and
we have assumed that the derivatives of $\phi_k$ do not depend on the axial variable $x$.
For an effectively 1D system, we assume that the transverse wave function remains in its Gaussian ground-state,
$\phi_k=\frac{1}{\pi\sqrt{\sigma_k}}\exp(-\frac{r_{\perp}^2}{2\sigma_k^2})$.
To account for axial effects, we allow the width $\sigma_k$ to be a
variational parameter, $\sigma_k=\sigma_k[f_k(x,t)]$; this yields:
%
%
\begin{equation}
\mu_{\perp k}= \frac{\hbar^2}{2M} \sigma_k^{-2}
+\frac{M}{2}\omega_{\perp}^2 \sigma_k^2
+\frac{g_{kk}|f_k|^2}{2\pi} \sigma_k^{-2}
+\frac{g_{12}|f_{3-k}|^2}{\pi (\sigma_1^2+\sigma_2^2)}.
\notag
\end{equation}
%
There are two different approaches
to determine $\sigma_k$: one can minimize the chemical potential $\mu_{\perp k}$
with respect to $\sigma_k$ for given $f_k(x,t)$ or, alternatively, one can use the Euler-Lagrange
equations from the Lagrangian associated to Eq.~(\ref{eq: f}), minimizing the total energy~\cite{Salasnich06}.
These two approaches lead to the following expression for $\sigma_k$,
\begin{equation}
 \sigma_k^4 =  \frac{\hbar^2}{\omega_{\perp}^2 M^2}
 + \frac{g_{kk}|f_k|^2}{A\pi M \omega_{\perp}^2}
 + \frac{2g_{12}|f_{3-k}|^2}{\pi M \omega_{\perp}^2 (\sigma_1^2+\sigma_2^2)^2}\sigma^4_k,
\label{eq: sigma}
\end{equation}
where parameter $A=1$ corresponds to the GMDE system
and $A=2$ for the NPSE system.
Notice that Eqs.~(\ref{eq: f}) and (\ref{eq: sigma}) constitute a
set of coupled nonlinear equations which have to be solved consistently in order to obtain $f_k(x,t)$ and $\sigma_k[f_k(x,t)]$.

Using the above approach, we
will investigate the
trapped dynamics of a DB soliton in a quasi-1D condensate.
For the
1D case without a trap in the axial direction ($V(x)=0$), and assuming that
all scattering lengths are equal, there exists an
analytical DB soliton solution of Eqs.~(\ref{eq: coupled GP 3d})~\cite{buschanglin};
this can be expressed in the following dimensionless form (in units so that $\hbar=M=1$),
\begin{eqnarray}
\psi_D &=& \imath \sqrt{\mu} \sin \alpha
+ \sqrt{\mu}\cos\alpha \tanh(\kappa(x-q(t))),
\label{eq: dark soliton}
\\[0.5ex]
\psi_B &=& \sqrt{\frac{N_B\kappa}{2}} e^{\imath(\phi+\omega_B t
+x\kappa\tan \alpha)}{\rm sech}(\kappa(x-q(t))).
 \label{eq: bright soliton}
\end{eqnarray}
Here, $\psi_D$ is the dark soliton (on top of a constant background
with chemical potential $\mu=\mu_D$),
with an inverse width $\kappa=\sqrt{\mu\cos^2\alpha+(N_B/4)^2}-N_B/4$, position
$q(t)=q(0)+t\kappa\tan\alpha$ and phase angle $\alpha$,
whereas
$\psi_B$ is the bright soliton that is symbiotically supported by the
dark one with the same width and position.
%
\begin{figure}[tbp]
\begin{center}
\includegraphics[width=8.5cm]{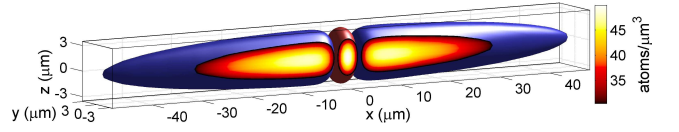} 
\end{center}
\vskip-0.4cm
\caption{(Color online)
Iso-level contours at 2/5 of the
maximal density (dark/bright soliton depicted in blue/red) of a
DB soliton as a result of 3D GPE simulations ($N_D=93\,367$, $N_B=7\,926$).
The transverse cut ($y=$const.) shows the atom density with the scale depicted
by the colorbar.
}
\label{fig1}
\end{figure}
%
In the realistic case of the hyperfine
states $|1,-1\rangle$ and $|2,0\rangle$ of $^{87}$Rb,
the scattering lengths
are different ($a_{11} = 100.86 a_0$, $a_{22} =  94.57 a_0$ and $a_{12} =  98.98 a_0$).
Nevertheless, in
the quasi-1D setting (with the trap), we have found that there exists
a stationary DB state [cf. Eqs.~(\ref{eq: dark soliton})-(\ref{eq: bright soliton}) with $\alpha=0$]
located at the trap center.
We identify this
state, $f^{\mathrm{stat}}_k$, using a fixed-point algorithm, and then perform a Bogoliubov-de-Gennes (BdG) analysis to determine its linear stability
by using the ansatz
$
f_k=f^{\mathrm{stat}}_k+ \left(u_k(x)\exp(\imath\omega t)
 + v_k^{\star}(x)\exp(-\imath\omega^{\star} t)\right).
$
The eigenfrequencies $\omega$ and amplitudes ($u_k$,$v_k$)
of the ensuing BdG linearization operator encode the dynamical stability of the system: for vanishing imaginary part
$\omega_i$ of $\omega=\omega_r + \imath \omega_i$, the system is dynamically
stable and $\omega_i \neq 0$ implies dynamical instability.
We also note that the eigenfrequency of the anomalous (negative energy) mode of the spectrum (see below)
coincides with the oscillation frequency of the DB soliton, similarly to dark solitons in one-species BECs~\cite{ourstuff}.
\begin{figure}[htb!]
\begin{center}
\includegraphics[width=8.0cm]{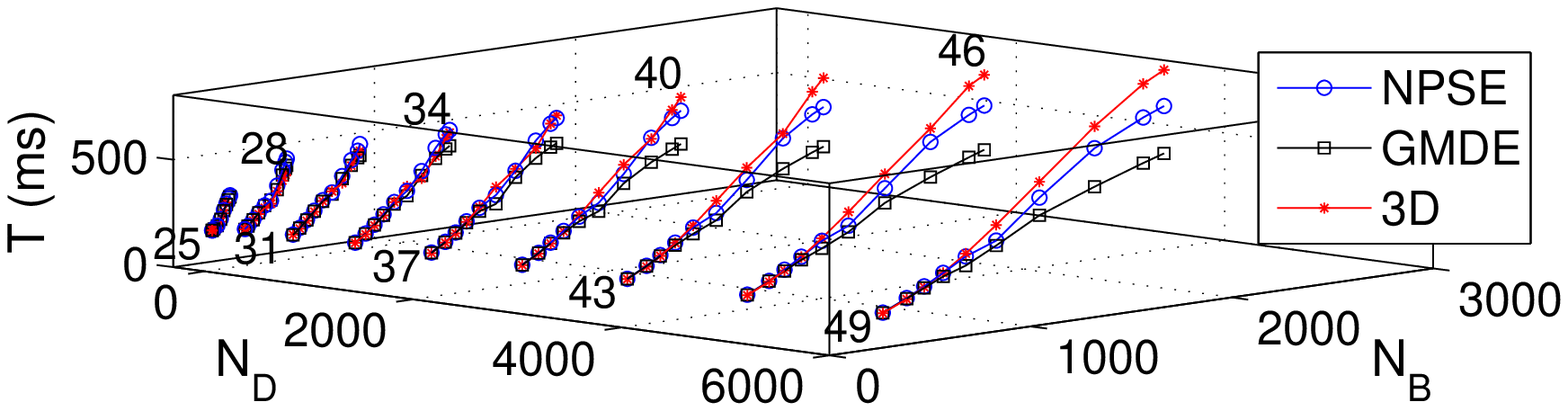} 
\end{center}
\vspace{-0.8cm}
\begin{center}
\includegraphics[width=5.0cm,height=8cm,angle=270]{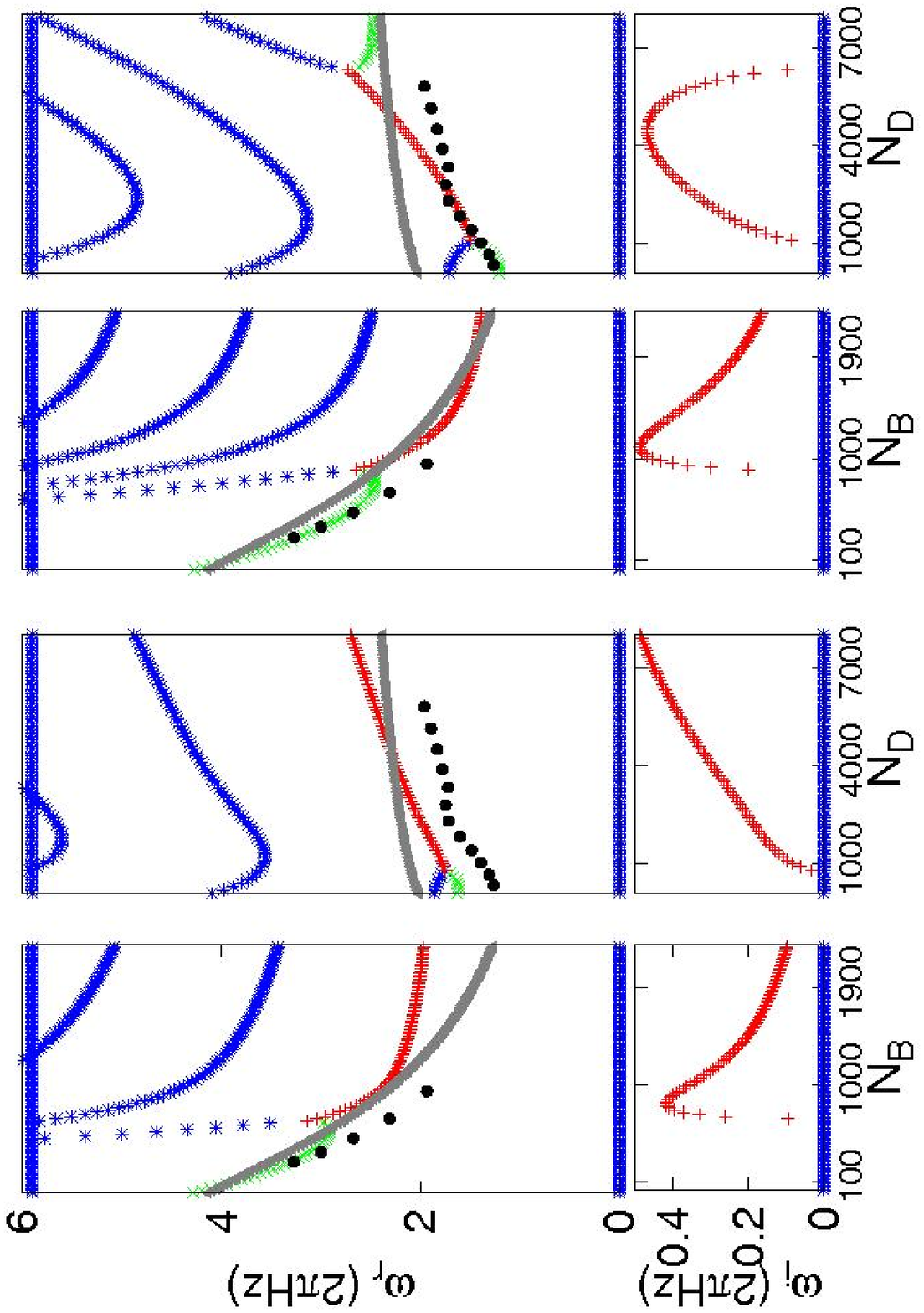}
\end{center}
\vskip-0.4cm
\caption{(Color online)
Top: DB soliton oscillation period vs.~$N_D$ and $N_B$
for fixed $\mu_D$ indicated by the numbers in the panel.
Circles (blue) and  squares (black)
depict the results from the BdG analysis of the GMDE
and NPSE, respectively; small (red)
dots represent results of
the 3D GPE.
Bottom: BdG analysis of GMDE (left) and NPSE (right) [real $\omega_r$ and
imaginary $\omega_i$ parts of the eigenfrequencies]: Anomalous mode (green),
dynamically unstable mode (red), analytical 1D result of
Ref.~\cite{buschanglin} (gray), and 3D GPE results (black points).}
\label{fig2}
\end{figure}

{\it Results.} We have chosen a cylindrical trap with frequencies
$\omega_\perp=2\pi\times133$ Hz and $\omega_x=2\pi\times5.9$ Hz,
similar to the
ones used in the experiment of Ref.~\cite{hamburg}.
In Fig.~\ref{fig1} we show iso-level contours of a DB
soliton resulting from numerical integration
of the 3D GPE, while
in Fig.~\ref{fig2} we compare the oscillation period (frequency) derived by
the effective 1D model against results of the 3D GPE.
The top panel illustrates the dependence of the period of the DB soliton on the number of atoms $(N_D,N_B)$ of the two components. It is clear that variation of the number of atoms, especially in the bright component by a factor of $2$, may lead to a significant (approximately two-fold) variation
of the DB soliton frequency. The agreement between 1D and 3D generally becomes worse when $N_B$ is increased and, to a lesser extent, when $N_D$ is increased. Notice that the NPSE model yields generally
more accurate predictions than the GMDE one.

The bottom panels of Fig.~\ref{fig2} show the DB soliton spectrum.
When the anomalous mode collides with a mode of positive energy, the
DB soliton
becomes dynamically unstable, i.e., the amplitude of its oscillation increases.
Such collisions are present in the DB soliton spectrum, and denote a critical difference in comparison with the case of
dark solitons in single-species BECs.
Furthermore, as observed in the right-most panel of Fig.~\ref{fig2}
and also corroborated by our 3D GPE simulations, this oscillatory instability disappears for sufficiently large $N_D$.

From Fig.~\ref{fig2}, we infer that for progressively larger atomic
populations, a departure from the quasi-1D behavior emerges and a
larger (smaller) oscillation frequency is obtained, if $N_D$ ($N_B$) is increased.
\begin{figure}[htb!]
\begin{center}
\includegraphics[width=2.8cm]{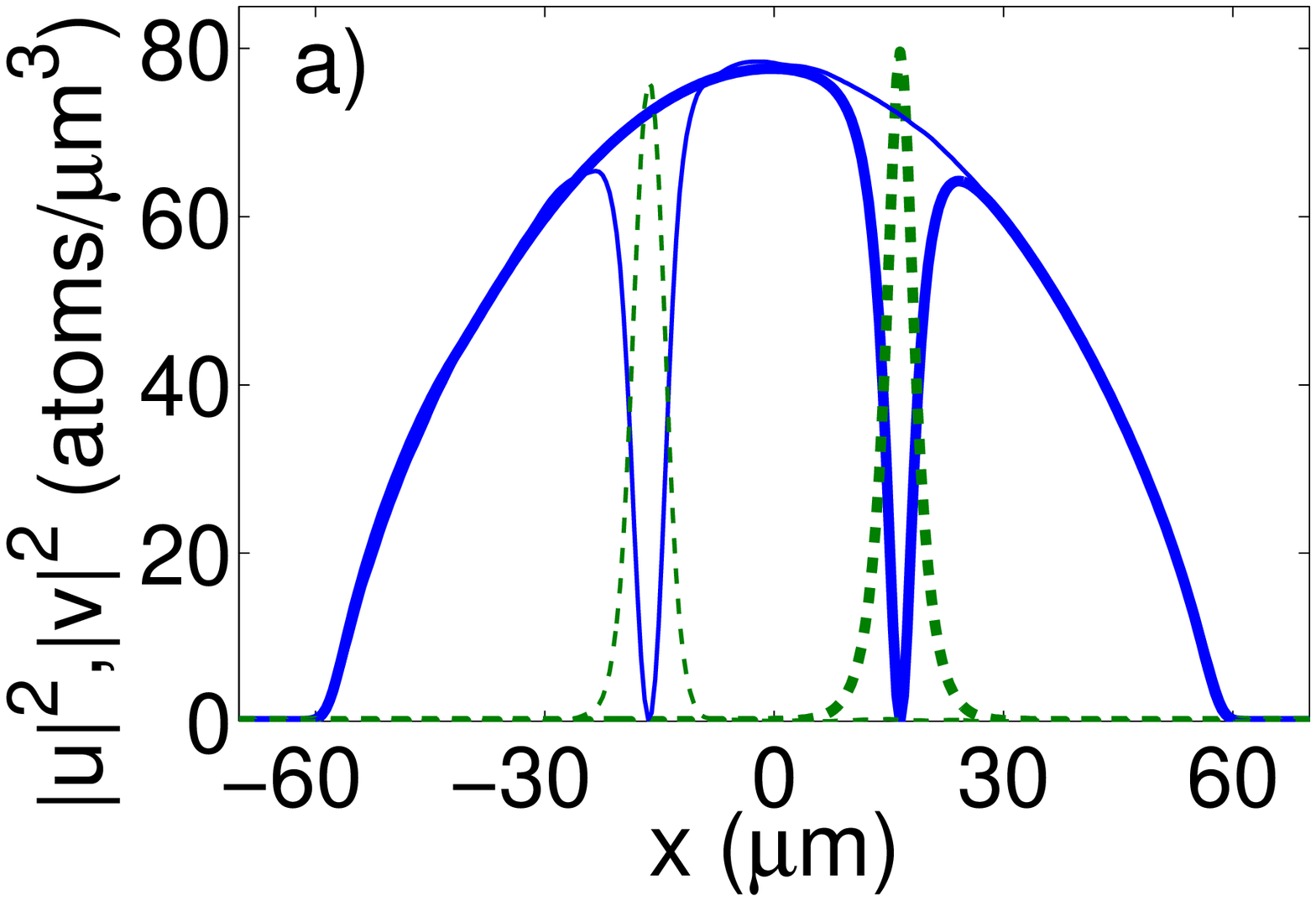} 
\includegraphics[width=2.8cm]{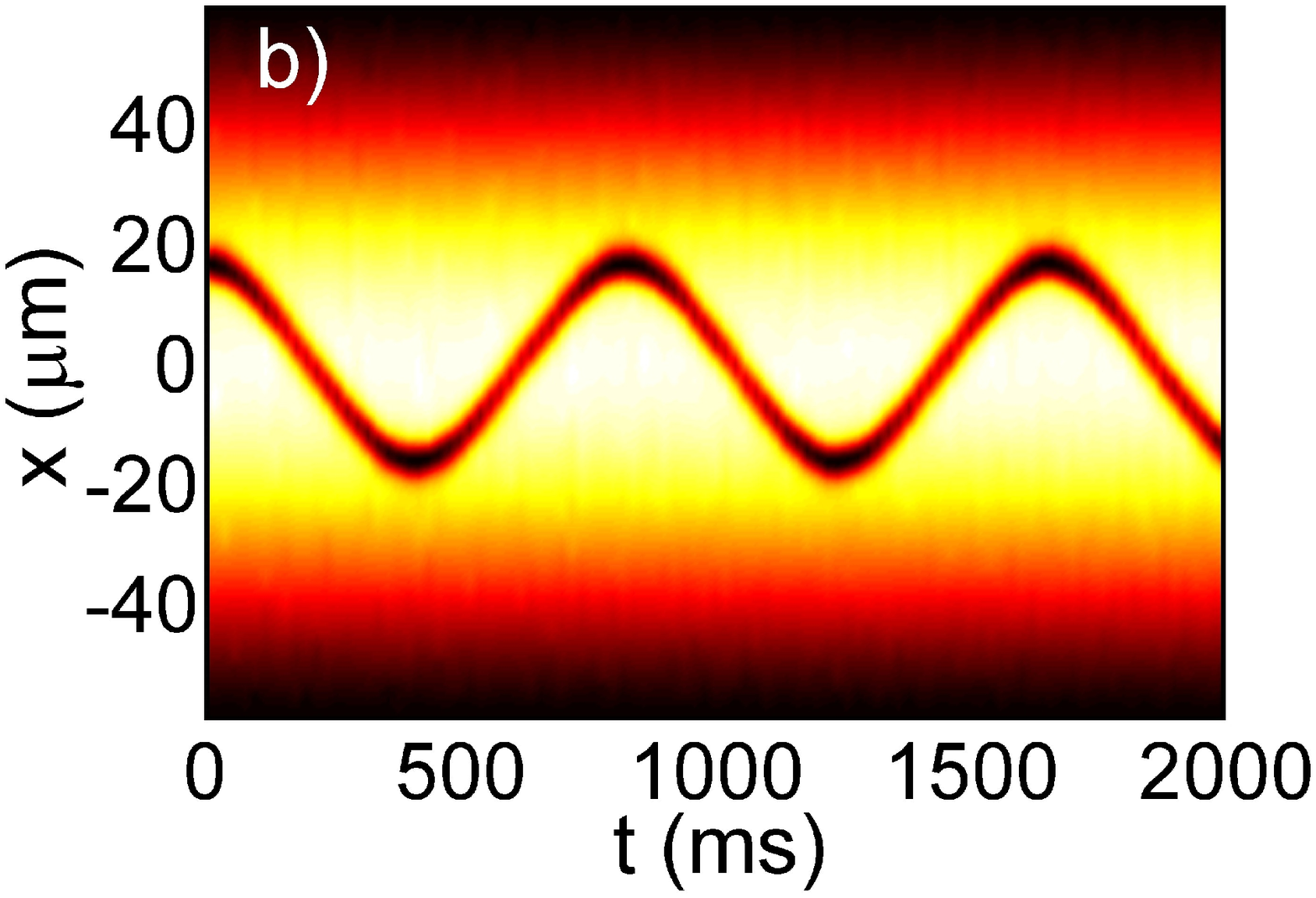} 
\includegraphics[width=2.8cm]{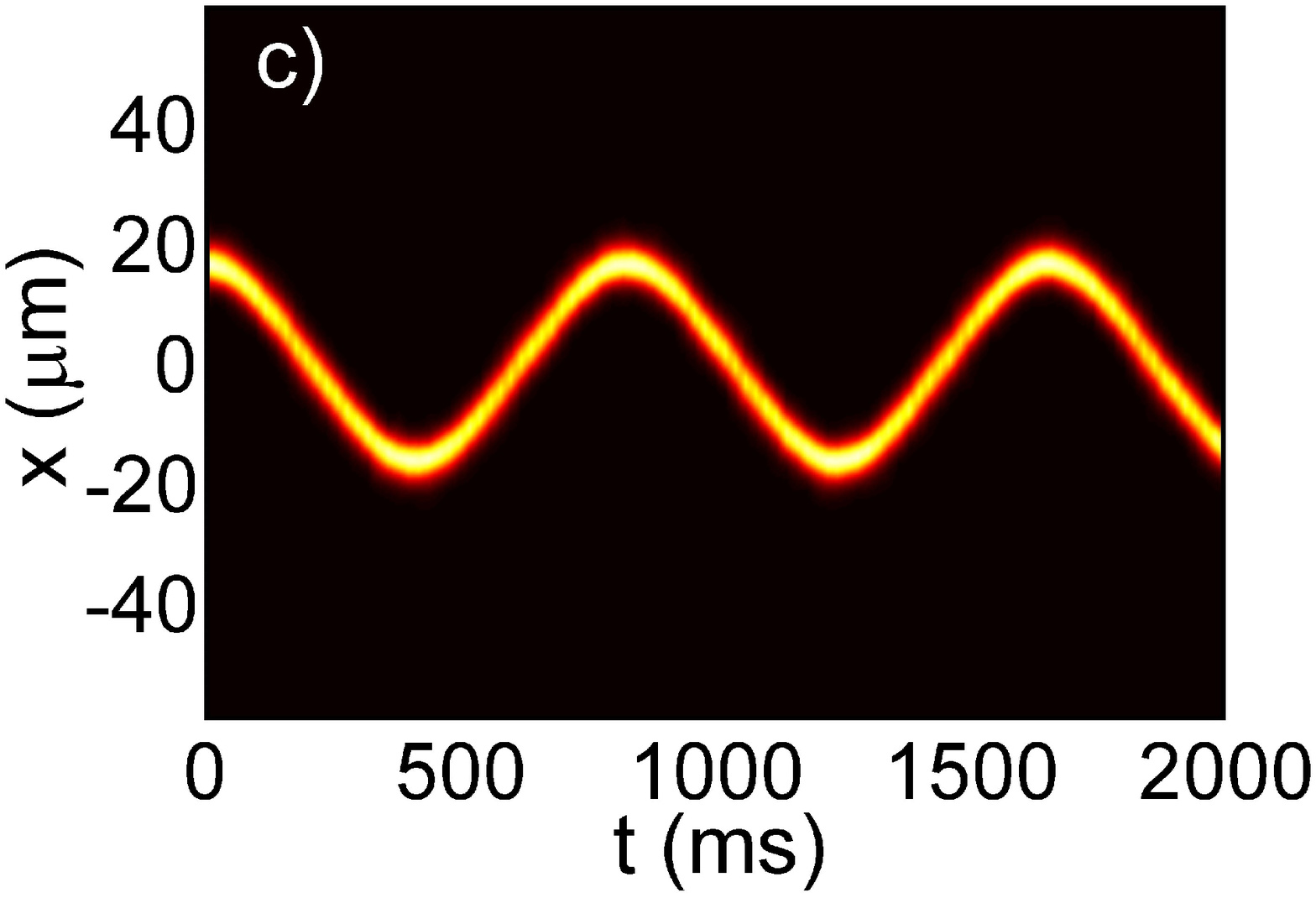} 
\end{center}
\vspace{-0.5cm}
\caption{(Color online)
An oscillating DB soliton with period $T=827$ ms in the 3D GPE approach.
Left panel: Transversal ($y=z=0$) cut of
the density at $t=0$ (thick lines) and
$t=T/2$ (thin lines). Solid (dashed) line depicts
the density for the dark (bright) component.
Middle/right panels: Contour plots showing the evolution of the
$(y,z)$-integrated density of the dark and bright solitons.
}
\label{fig3}
\end{figure}

According to Fig.~\ref{fig2}, the atom numbers used in Ref.~\cite{hamburg} (up to $N_B\sim8000$ and $N_D\sim92000$)
are out of the realm of validity of the effective 1D equations; thus, in this case, the 3D GPE has to be applied.
Figure \ref{fig3} illustrates the oscillating DB soliton for trap frequencies and atom numbers comparable
to those used in Ref.~\cite{hamburg}. This 3D simulation results in an oscillating DB period of about 827 ms
that is comparable with the period observed in Ref.~\cite{hamburg} (slightly larger than 1 s).
Possible sources for the
discrepancy between the numerical and experimental oscillation periods include
(i) high sensitivity of the period on the $N_D$ to $N_B$ ratio,
(ii) sensitivity of the dynamics to uncertainties in the measured scattering lengths,
and, perhaps more importantly,
(ii) our numerics confirm that the oscillations are not harmonic and tend to have increasing
periods for higher oscillating amplitudes.

\begin{figure}[tbp]
\begin{center}
\includegraphics[width=8cm]{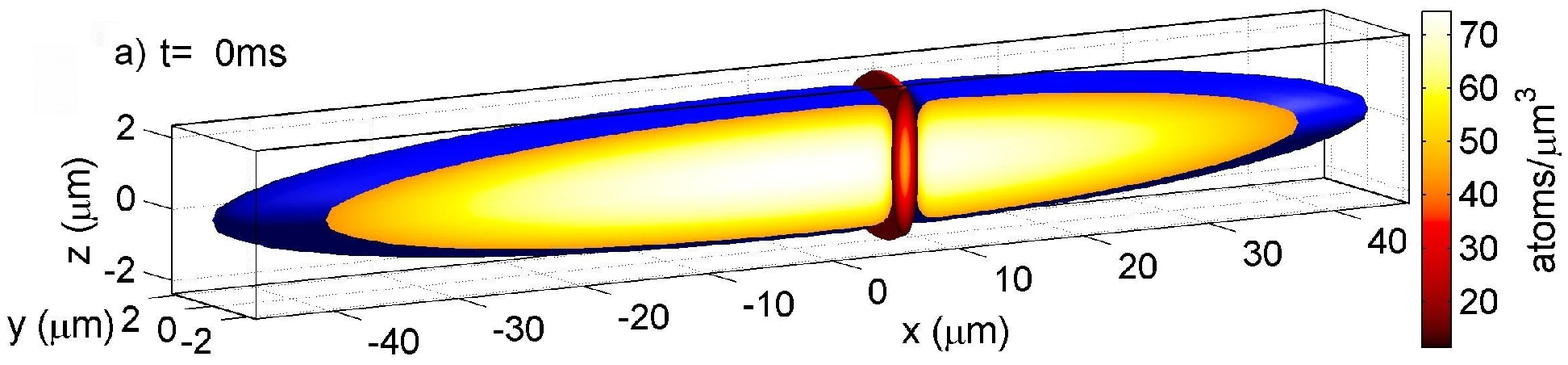} 
\includegraphics[width=8cm]{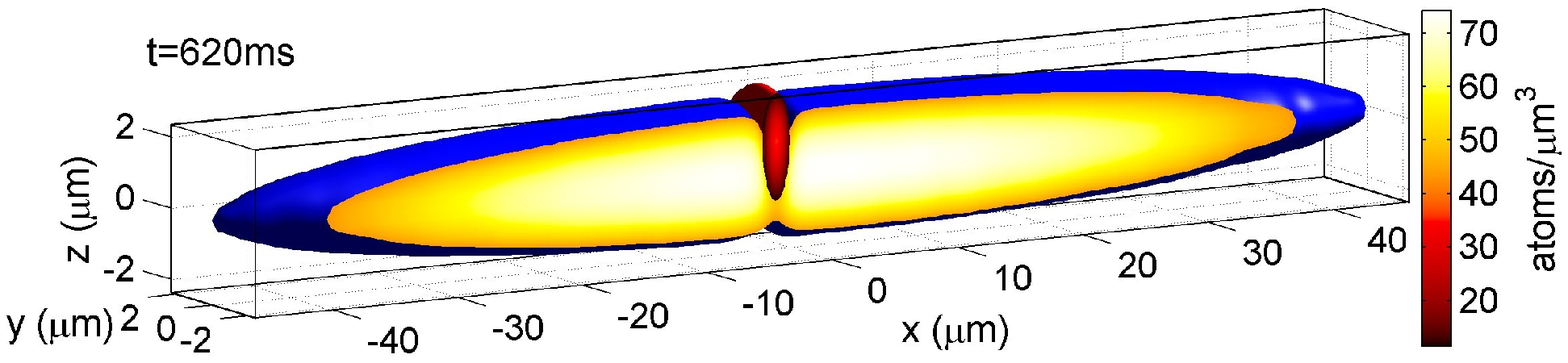} 
\\[1.5ex]
\includegraphics[width=7cm]{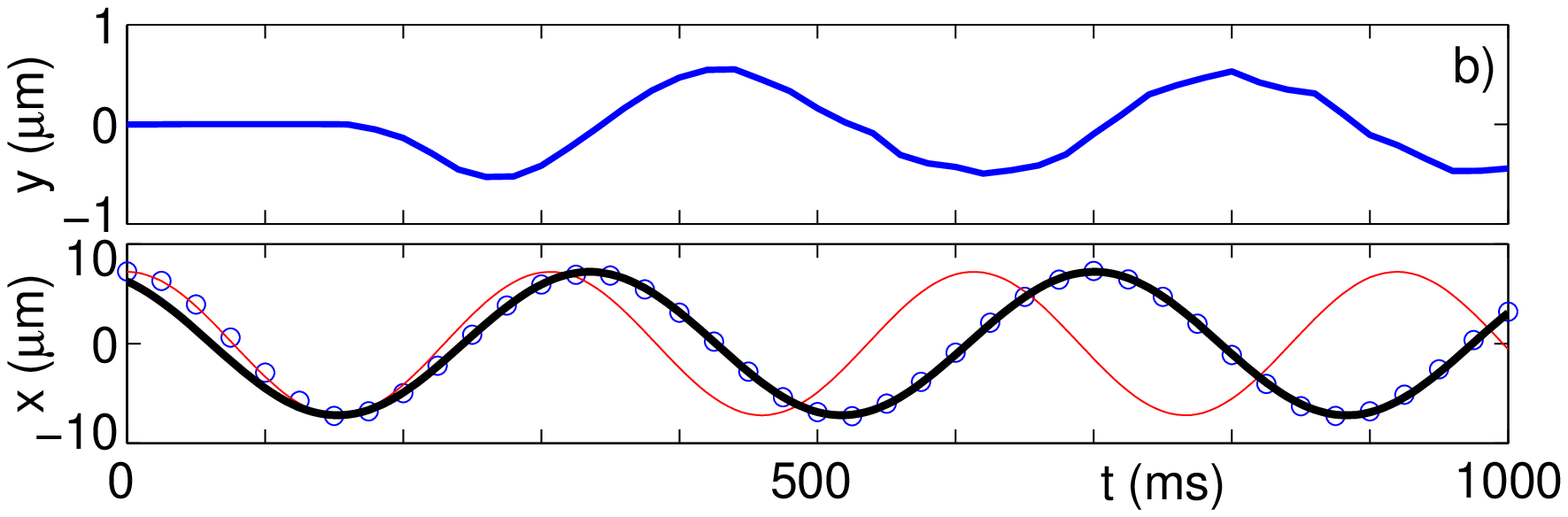} 
\\[1.5ex]
\includegraphics[width=2.92cm,height=2.2cm]{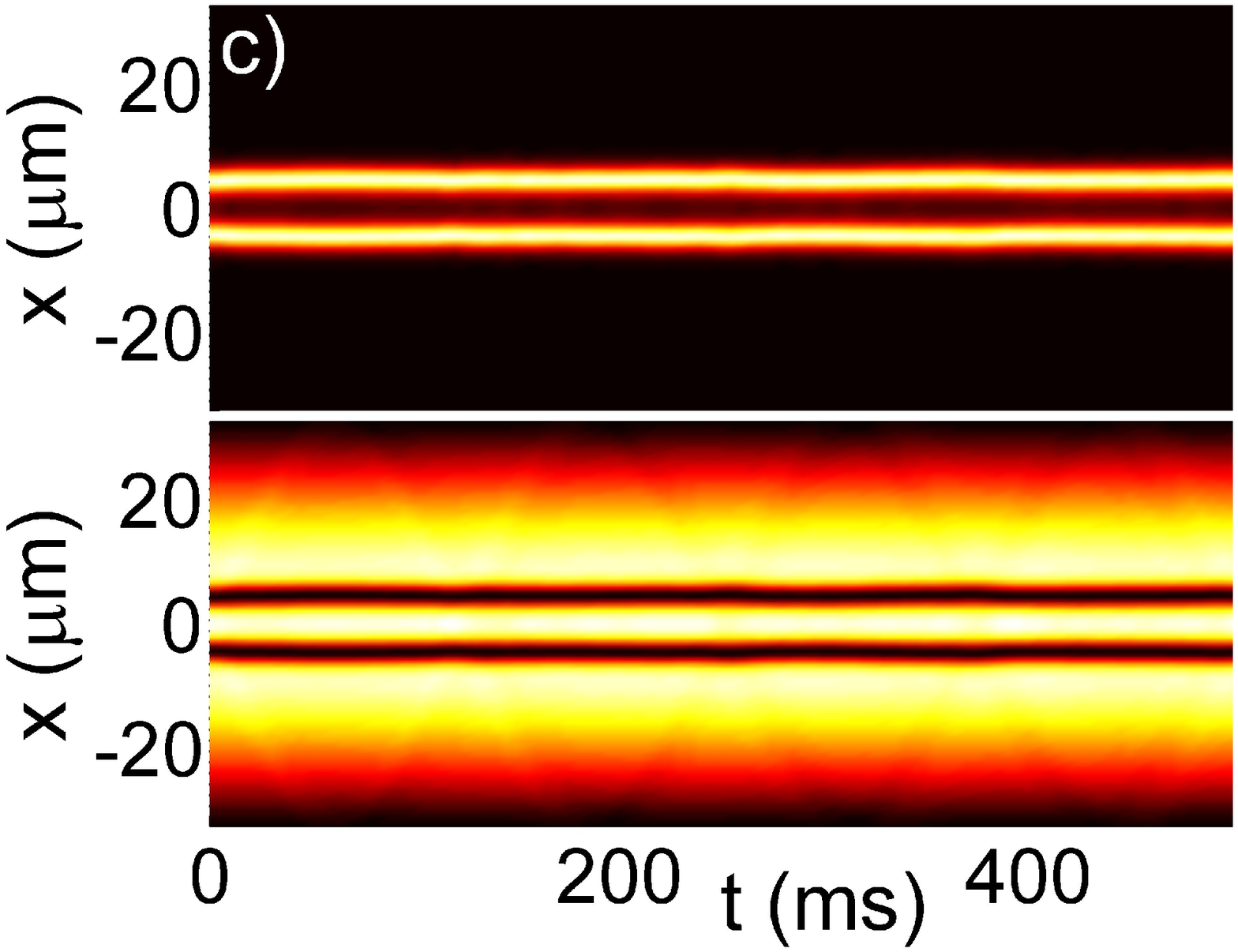} 
\includegraphics[width=2.50cm,height=2.2cm]{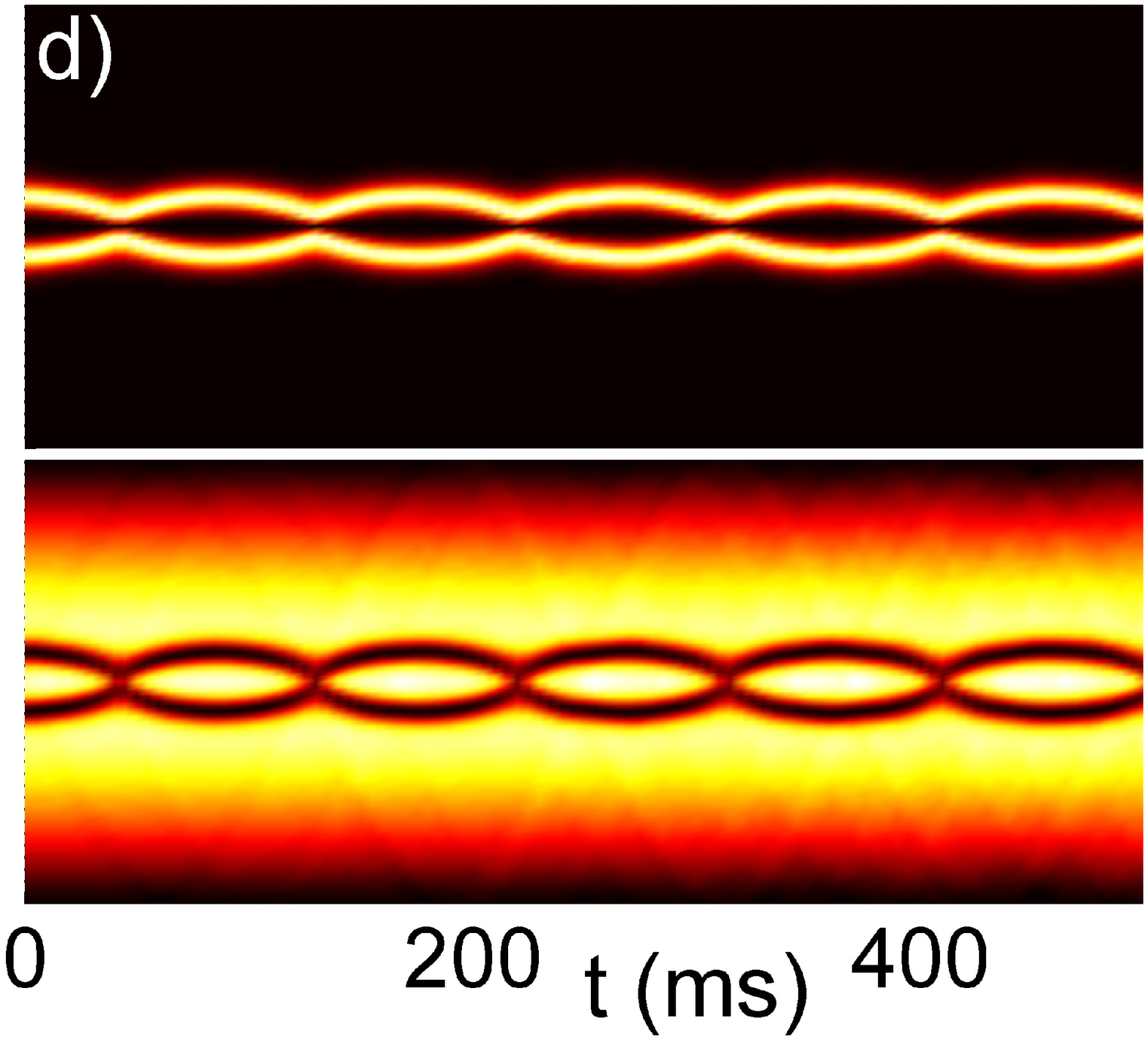} 
\includegraphics[width=2.50cm,height=2.2cm]{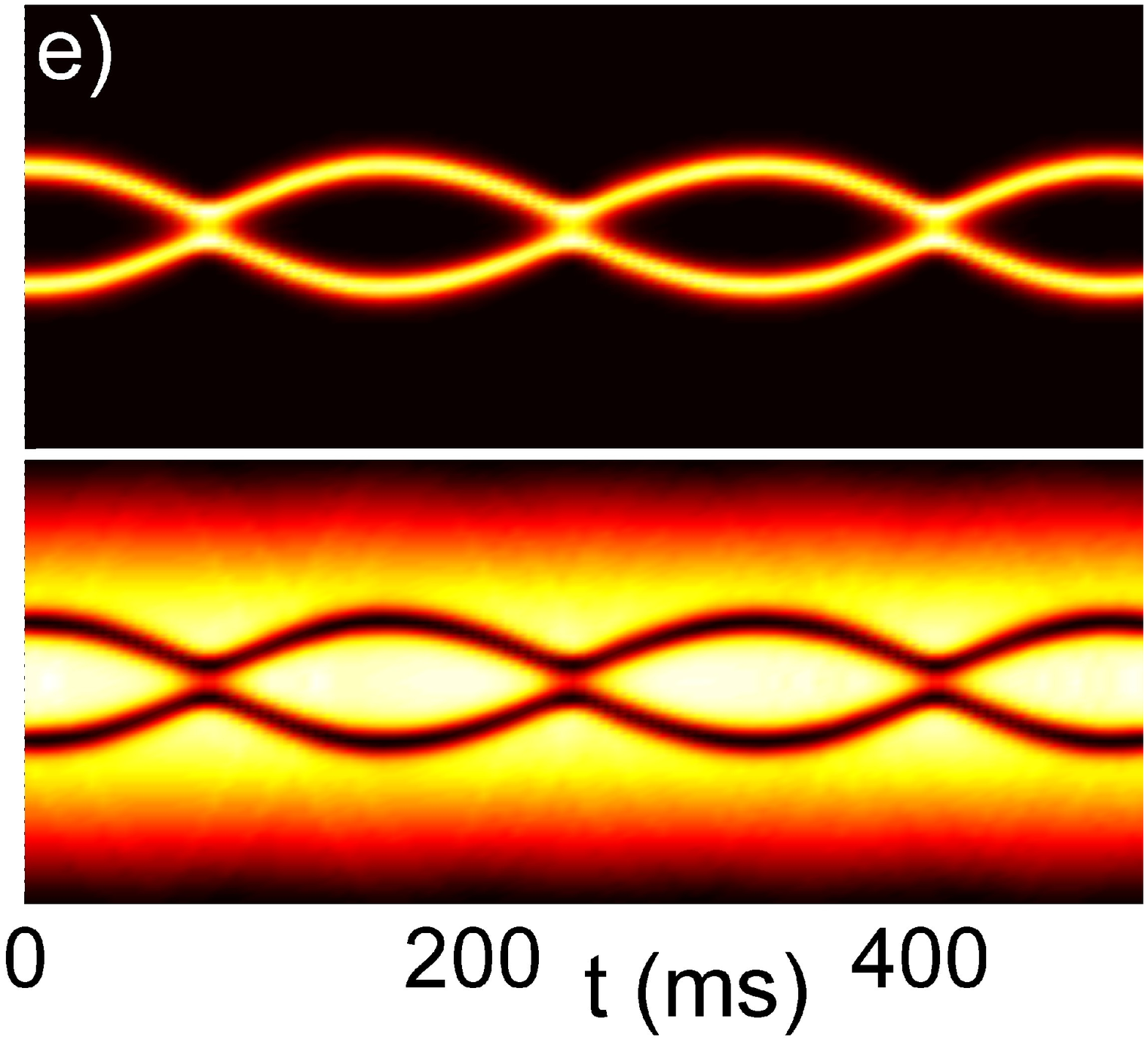} 
\end{center}
\vskip-0.4cm
\caption{(Color online)
%
(a) Transverse oscillations of the DB soliton
for $(N_D,N_B)=(88\,181,1\,058)$. Top subpanel: initial condition. Bottom subpanel: a snapshot of the oscillating DB soliton at $t=620$ ms.
(b) Transverse (top) and longitudinal (bottom) oscillations for the
bright soliton. The thin (red) and thick
(black) solid lines are sinusoidal fits to
the longitudinal oscillations (circles) yielding periods $T=306$ ms (for
$t<200$ ms) and $T=365$ ms (for $t>200$ ms), respectively.
Bottom row of panels: Interaction of {\em two} DB
solitons [top (bottom) subpanels depicting the bright (dark)
component].
(c) In-phase (i.e.~repulsive among bright) solitons close to their equilibrium position.
(d) Out-of-phase (i.e.~attractive among bright) solitons starting at the same
location as in panel c).
(e) Multiple collisions of in-phase solitons.
}
\label{fig4}
\end{figure}

The departure from the effective 1D description can be noticed, e.g.,~in panels (a) and (b) of Fig.~\ref{fig4}.
It is clear both from the oscillation snapshots (Fig.~\ref{fig4}(a)) and from the evolution of the bright soliton center
in the transverse direction (Fig.~\ref{fig4}(b) top panel), that the DB soliton starts exploiting the transverse degrees
of freedom shortly after release. Up to $t<200$~ms the soliton is at rest with respect to the transverse direction.
For $t>200$~ms an oscillation of the soliton occurs in the transversal direction leading at the same time to a
reduction of the oscillation frequency in the axial direction (cf. thick (black) solid line in the bottom panel of
Fig.~\ref{fig4}b).

We also showcase 3D GPE simulations, cf.~bottom row of
panels in Fig.~\ref{fig4}, depicting the interaction between {\em two}
DB solitons \cite{berloff}. Panel (c) shows a stationary DB pair with
{\em in-phase}, i.e.~mutually repulsive, bright soliton components
(also, the dark ones always repel)
that is balanced by the pull of the harmonic trap.
Panel (d) depicts the evolution of the same initial DB pair as in
panel (c), but with {\em out-of-phase}, i.e.~mutually attractive,
bright solitons, which yields an oscillatory dynamics.
Panel (e) depicts the oscillations and collisions for in-phase
bright solitons that were released at larger distances from the trap center and thus cannot avoid colliding despite their mutual repulsion.
It is noteworthy that the DB collisions are apparently nearly elastic
as the DB solitons retain their shape even after multiple
collisions.

\begin{figure}[tbp]
\begin{center}
~\includegraphics[width=8.5cm]{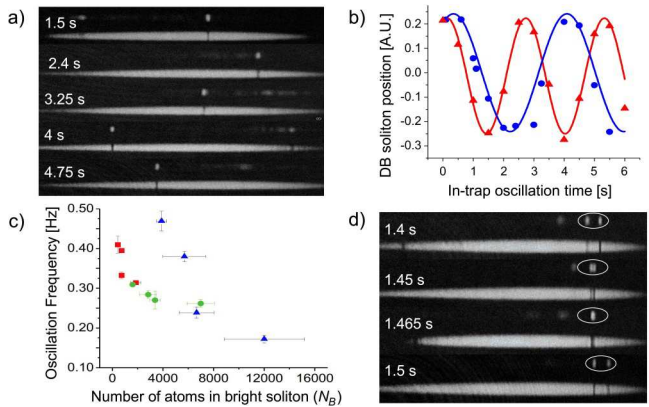}
\end{center}
\vskip-0.4cm
\caption{(Color online)
(a) Experimental images showing the oscillation of the DB soliton at
the in-trap evolution times indicated.
(b) In-trap DB soliton oscillation.  Triangles (red) and squares (blue)
correspond to, respectively,
$(N_B,N_D) \approx (680,27\,000)$ and
$(N_B,N_D) \approx (9\,000,650\,000)$. Solid lines correspond to
fitted harmonic oscillations with frequencies $0.39$ Hz and $0.27$ Hz
respectively.
(c) Oscillation frequency vs.~$N_B$ for different number of atoms in the
dark component:
squares (red): $N_D \approx 30\,000$,
triangles (green): $N_D \approx 200\,000$, and
circles (blue): $N_D \approx 430\,000$.
(d) Experimental expansion images showing the oscillation and
collision (in circled regions) of two DB solitons.
%
Images in (a) and (d) are taken after 7~ms and 8~ms of free expansion for the bright and dark component, respectively.
The components are vertically overlapped prior to expansion.
}
\label{fig_exp}
\end{figure}

Finally, we present experimental data corroborating some principal points of
our analysis (Fig.~\ref{fig_exp}).
Nonlinear effects in the counterflow of two BEC components are exploited to generate individual DB solitons~\cite{Hamner}.
The dark and bright component are formed by $^{87}$Rb atoms in the $|1,-1 \rangle$ and
$|2,-2\rangle$ state, respectively, for which $a_{11} = 100.4~a_0$, $a_{22} =  98.98~a_0$,
and $a_{12} =  98.98~a_0$~\cite{Kokkelmans}. The atoms are held in an elongated optical dipole trap with
trapping frequencies $\omega_{x,y,z} =2\pi\times \{1.3,163,116\}$~Hz. While the lack of exact cylindrical
symmetry as well as the large atom number in the experiment preclude a direct comparison with our analytic results, the
experiment clearly shows the anticipated decrease of oscillation frequency with increasing number of atoms in the bright
and decreasing number of atoms in the dark component. Experimental results of the collision between two DB solitons are presented in Fig.~\ref{fig_exp}(d),
confirming their near-elastic nature.

{\it Conclusions.}
We characterized the effectively 1D dynamics of DB solitons and showcased
their potential dynamical instability.
We demonstrated experimentally and theoretically the tunability of the
oscillation frequency of a DB soliton.
A spontaneous breaking of the cylindrical symmetry resulting
in a reduction of the DB oscillation frequency
was
predicted, along with (also observed) near-elastic collisions, as well as a
strong phase dependence of the collisional dynamics of DB solitons.
Future directions include
a detailed effective particle-based understanding of the DB soliton
interactions,
as well as a generalization of this picture towards the precession
and interactions of vortex-bright solitons.

\end{document}